\documentclass[
twocolumn,
superscriptaddress,
longbibliography,
pra,
tightenlines,
aps,
]{revtex4-1}
\pdfoutput=1

\usepackage[utf8]{inputenc}
\usepackage[T1]{fontenc}
\usepackage[english]{babel}
\usepackage[normalem]{ulem}
\usepackage{amsmath}
\usepackage{amssymb}
\usepackage{amsfonts}
\usepackage{mathrsfs}
\usepackage{dsfont}
\usepackage{bm} 
\usepackage{textgreek} 
\usepackage{dcolumn}
\usepackage{graphicx}
\usepackage[caption=false]{subfig}
\usepackage{enumerate}
\usepackage[shortlabels]{enumitem}
\usepackage{hyphenat}
\usepackage[bookmarks=false,colorlinks=true,linkcolor=blue,citecolor=blue,urlcolor=blue]{hyperref}
\usepackage[capitalise]{cleveref}
\crefname{section}{Sec.}{Secs.}
\Crefname{section}{Section}{Sections}
\usepackage{hypcap} 
\usepackage{floatrow}
\usepackage{tablefootnote}
\usepackage{array}
\usepackage{braket}
\usepackage{siunitx}

\begin{document}

\title{Experimental quantum voting using photonic GHZ states}

\author{F.~Joseph~Marcellino}
\affiliation{Department of Applied Physics, University of Geneva, CH-1211 Geneva, Switzerland}

\author{Mingsong Wu}
\affiliation{Department of Applied Physics, University of Geneva, CH-1211 Geneva, Switzerland}

\author{Rob~Thew}
\email[Corresponding author: ]{Robert.Thew@unige.ch}
\affiliation{Department of Applied Physics, University of Geneva, CH-1211 Geneva, Switzerland}

\date{\today}

\begin{abstract}
Quantum communication protocols seek to leverage the unique properties of quantum systems for coordination or communication tasks, usually with guarantees of security or anonymity that exceed what is possible classically. One promising domain of application is elections, where strong such guarantees are essential to ensure legitimacy. We experimentally implement a recently proposed election protocol from Centrone \textit{et al.} designed such that no one, including a potential central authority, can know the preferred candidate of any voter other than themself. We conduct a four-party election, generating and distributing four-partite GHZ states with $\approx 89\%$ fidelity and successfully recording voters' intentions $\approx 87\%$ of the time.

\end{abstract}

\maketitle

\textit{Introduction} -- Many quantum phenomena (e.g. no-cloning, true randomness, destructive measurement) are attractive for cryptographic or communication purposes, where they facilitate protocols that can outperform any classical counterpart, the most salient example being quantum key distribution, which allows for key distribution at a distance with information-theoretic security. Further advantages are offered by entangled quantum states, which exhibit correlations that cannot be reproduced classically. This fact enables such states to be verified even when shared between mutually distrustful parties~\cite{Pappa2012}, and is at the core of a variety of protocols for tasks including device-independent key distribution~\cite{Zapatero2023}, secret sharing~\cite{Hillery1999}, Byzantine agreement~\cite{Fitzi2001}, anonymous transmission~\cite{Christandl2005}, and secure distributed computation~\cite{Barral2024}.

Electronic elections are a natural candidate use case for quantum communication, as they necessitate communication between potentially distrustful or dishonest parties, and ideally should guarantee privacy and anonymity. Centrone \textit{et al.}~\cite{Centrone2022} propose a protocol based on n-partite GHZ states meeting these criteria; it does not require the voters to trust each other nor the source of the associated quantum states, is robust against (potentially colluding) dishonest participants, and is both provably correct (the intentions of the voters are faithfully recorded) and anonymous (no voter can be identified with their vote), avoiding the vulnerabilities~\cite{Arapinis2021} characteristic of earlier quantum voting schemes~\cite{Wang2016, Bonanome2011, Vaccaro2007, Okamoto2008}. The protocol incorporates a previously proposed entanglement verification scheme~\cite{Pappa2012}, which both ensures the integrity of the state and provides bounds on the probability of error/fraud.

We implement a version of this protocol, generating and distributing four-partite polarization-entangled photonic GHZ states. We achieve 89\% fidelity to the target state, resulting in successful voting/state verification $87\pm3\%$ of the time. This serves as a proof-of-principle for our modified protocol, which has the advantage of being fully implementable with current technology and requiring no additional assumptions on the behavior of the source.

\begin{figure}
\centering
\includegraphics[width = 0.95\textwidth]{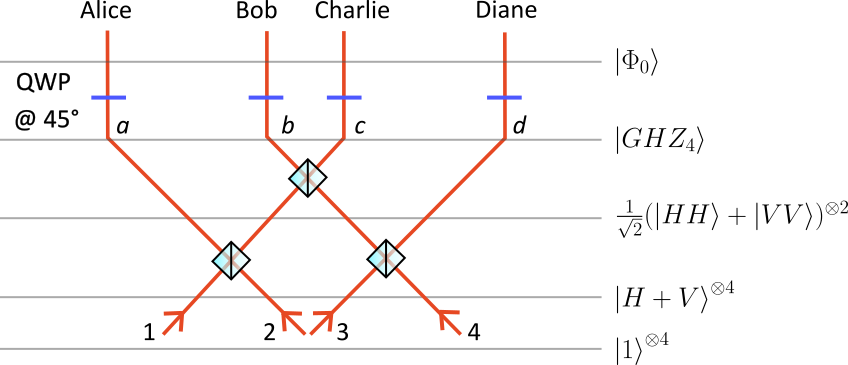}
\caption{\label{fig:fig_schematic} Schematic representation of state generation scheme. The quantum state after fourfold postselection is indicated on the right. QWP, quarter waveplate.}
\end{figure}

\textit{State generation} --  To generate the desired four-photon state for this protocol, we start with four single photons, as shown in Fig.~\ref{fig:fig_schematic}. We rotate all four photons to diagonal polarization, thus beginning with the state, 
\begin{equation}
    \begin{split}
        \ket{\Psi}_\mathrm{in} = &\frac{1}{4}(\ket{H}_{1} + \ket{V}_{1})\otimes(\ket{V}_{2} + \ket{H}_{2})\otimes \\
        & (\ket{H}_{3} + \ket{V}_{3})\otimes(\ket{V}_{4} + \ket{H}_{4})
    \end{split}
\end{equation}
Photons $1$ and $2$ are combined on one PBS, and similarly, $3$ and $4$ on another PBS. One output mode of each PBS is then sent to a third PBS, such that we distribute a four-partite entangled state to Alice, Bob, Charlie and Diane. The reduced state after fourfold postselection is then the four-photon GHZ state
\begin{equation}
    \ket{\Psi}_\mathrm{out} = \frac{1}{\sqrt{2}}(\ket{H}_{a}\ket{H}_{b}\ket{H}_{c}\ket{H}_{d} + 
    \ket{V}_{a}\ket{V}_{b}\ket{V}_{c}\ket{V}_{d})
\end{equation}

To implement the voting protocol, we require instead the locally equivalent state
\begin{equation}
    \ket{\Phi_0} = \frac{1}{\sqrt{8}}(\sum_{\Delta(y)=0 \mod{4}}\ket{y} - \sum_{\Delta(y)=2 \mod{4}}\ket{y})
\end{equation}
where $\Delta(y)$ is the Hamming weight of the classical bitstring $y$, obtained by applying a Hadamard and $\sqrt{Z}$ gate to each photon (realized using a quarter waveplate at $45^{\circ}$). 

We note the following useful property of $\ket{\Phi_0}$: upon application of an operator consisting of 0 mod 4 Hadamard gates to a subset of the qubits (in the four qubit case, this means all or none, although the property holds for higher-dimensional analogues of the state) and $Z$ gates on the remaining qubits, $\ket{\Phi_0}$ goes to $\pm\ket{\Phi_0}$. Moreover, defining the orthogonal state $\ket{\Phi_1}$ as 

\begin{equation}
    \ket{\Phi_1} = \frac{1}{\sqrt{8}}(\sum_{\Delta(y)=1 \mod{4}}\ket{y} - \sum_{\Delta(y)=3 \mod{4}}\ket{y})
\end{equation}

it can be shown that, upon application of an operator consisting of 2 mod 4 Hadamard gates to a subset of the qubits and $Z$ gates on the remaining qubits, $\ket{\Phi_0}$ goes to $\pm\ket{\Phi_1}$.


\textit{Protocol} -- The original protocol in~\cite{Centrone2022} explicitly requires that the decision whether to verify the state or use it for voting be made after the state is generated and distributed, but before it is measured. This requires either quantum memories or the ability to decide and adapt the measurement appropriately on timescales comparable to the time it takes the state to travel from the source to the agents. We modify the original protocol to eliminate this requirement by adding random local basis choices and postselecting usable events. This comes at the expense of the following: we require more distributed states to execute the protocol, can no longer claim the same functional form for probability of anonymity as a function of probability of successful verification, and require at least three honest parties. A proof that the modified protocol guarantees anonymity in the limit that verification always succeeds is given in Appendix A.

Similar to quantum protocols for e.g. key distribution or Byzantine agreement, the core voting scheme itself is classical, with quantum resources used simply to distribute random but correlated bits. We first briefly describe the classical portion, which is identical to that of~\cite{Centrone2022}, then place it in context of our modified protocol.

Supposing $n$ voting agents and two candidates (``E'' and ``F''):

\begin{enumerate}
    \item Each agent is randomly assigned a unique, secret voter index $i$ via a secure classical protocol (``UniqueIndex'' in~\cite{Centrone2022}).
    \item For $j = 1:n$:
    \begin{enumerate}
        \item A source distributes a random, secret bit $b_{i_j}$ to each agent $i$ such that $\bigoplus_i b_{i_j} = 0$, i.e. the Hamming weight of the overall bitstring $b_{1_j}b_{2_j}...b_{n_j}$ is even (we refer henceforth to such bitstrings as ``parity-even'').  
        \item All agents except agent $i=j$ (the ``voter'') broadcast their secret bit. The voter broadcasts:
        \begin{itemize}
            \item their secret bit $b_{j_j}$ if they want to vote for Candidate E.
            \item the opposite of their secret bit $\neg b_{j_j}$ if they want to vote for Candidate F.
        \end{itemize}
        \item All broadcast bits are recorded as row $j$ of a public results board $R_{ji}$.
    \end{enumerate}
    \item $r_j = \bigoplus_i R_{ij}$ is computed for each row $j$, with $r_j = 0$ counting as a vote for Candidate E and $r_j = 1$ as a vote for Candidate F.
    \item Each voter $i$ anonymously indicates whether their vote was recorded successfully (i.e. whether $r_{j=i}$ matches their intentions) via a secure classical protocol (``LogicalOr'' in~\cite{Centrone2022}).
    \item If the proportion of indicated failures exceeds some allowed threshold, abort. Otherwise, declare the winner. 
\end{enumerate}

Note that this scheme already achieves voter anonymity (via encoding votes in an XOR), correctness, and robustness against malicious agents (via the confirmation step at the end), assuming that the source of bits in step 2a is trustworthy. The motivation for adding quantum resources is then to eliminate this assumption.

Recall that the task of the source is to distribute random sets of bits $\{b_i\}$ such that $\bigoplus_{i}b_i=0$. This can be achieved conveniently through the use of the $n$-qubit equivalent of $\ket{\Phi_0}$, which we term $\ket{\Phi_{0}^n}$, and upon which measurements in the computational or Hadamard basis yield results uniformly sampled from the space of parity-even $n$-bit strings. A source can also easily be verified to be distributing $\ket{\Phi_{0}^n}$ using only local rotations consisting of Hadamard and phase gates~\cite{Pappa2012}. We exploit these properties to enable the voting protocol for four voters as follows:

\begin{enumerate}
    \item [] For $k \in [1,l\gg 1]:$
    \item The source distributes state $\ket{\Psi}_k$ (nominally $\ket{\Phi_{0}^n}$ for all $k$) to the agents (each agent receives one photon per state).
    \item After receiving state $\ket{\Psi}_k$, each agent randomly chooses between measuring their photon in the computational basis, or measuring in the Hadamard basis (applying a Hadamard gate followed by a measurement in the computational basis). Each agent $i$ records the time of arrival, their chosen basis as $B_{i_k}$, and their outcome as $Y_{i_k}$. 
    \item [] After distributing $l$ states:
    \item The agents compare timestamps and postselect events where all four agents successfully received and measured a photon. For each fourfold event $p$:
    \begin{enumerate}
        \item The agents randomly choose one agent to act as the ``Verifier'' using a classical protocol (``RandomAgent'' in~\cite{Centrone2022}).
        \item All agents send their measurement basis $B_{i_p}$ to the Verifier.
        \item If the total number of Hadamard measurements is odd, the Verifier announces "discard," and the agents begin again from step (a) with the next event $p$. Else, the Verifier flips $m$ coins, where $m \gg 1$ is a security parameter.
        \item  If all return heads, the Verifier announces "voting," broadcasts $S_p \equiv \frac{1}{2}H_p$ mod 2, where $H_p$ is the total number of Hadamard measurements made on event $p$, and sets aside event $p$ for eventual voting.
        \item Else, the Verifier announces "verifying," and all agents send the Verifier their measurement result $Y_{i_p}$.
        \item The Verifier verifies that $\frac{1}{2}H_p=\sum_iY_{i_p}$ mod 2~\cite{Pappa2012}.
        \item If the event passes the test, the Verifier announces ``success'', else ``failure''.
    \end{enumerate}
    \item If the proportion of failed verifications exceeds some allowed threshold, abort. Else, use the events previously reserved for voting as described earlier, replacing the bits obtained in step 2a with the respective measurement outcomes, and taking the XOR with $S_p$ (such that rounds with $\frac{1}{2}H_p$ odd also give $\bigoplus_i R_{ij} = 0$ when the voter does not flip their bit).
\end{enumerate}

By postponing the decision whether to use a given state for voting or verifying until after the measurements have taken place, this version ensures that a malicious source cannot undermine the protocol by, for instance, distributing the target state during verification rounds and some other state during voting rounds, without requiring the ability to store states. Note that this is true even if the source is colluding with one or more of the agents. As alluded to earlier, one disadvantage is that, 50\% of the time, the agents will measure in a set of bases with an odd number of Hadamard measurements; these events are usable neither for verification nor voting. While unfortunate, this represents a purely linear increase in complexity that, given sufficiently high rates of state distribution, does not compromise the functionality of the protocol.

\begin{figure}
\centering
\includegraphics[width = 0.95\textwidth]{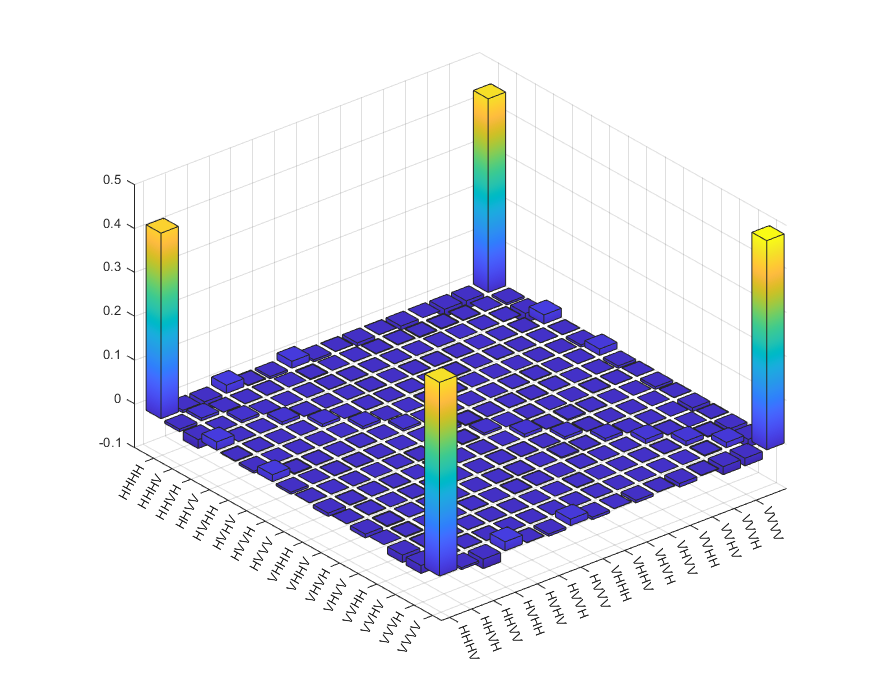}
\caption{\label{fig:fig_rho_real} Real part of the reconstructed four-qubit density matrix obtained from a quantum tomography measurement. Fidelity to the target state $F \approx .89$. All imaginary components have magnitude $\leq .02$.}
\end{figure}


\begin{figure*}[!t]
\centering
\includegraphics[width = 0.95\textwidth]{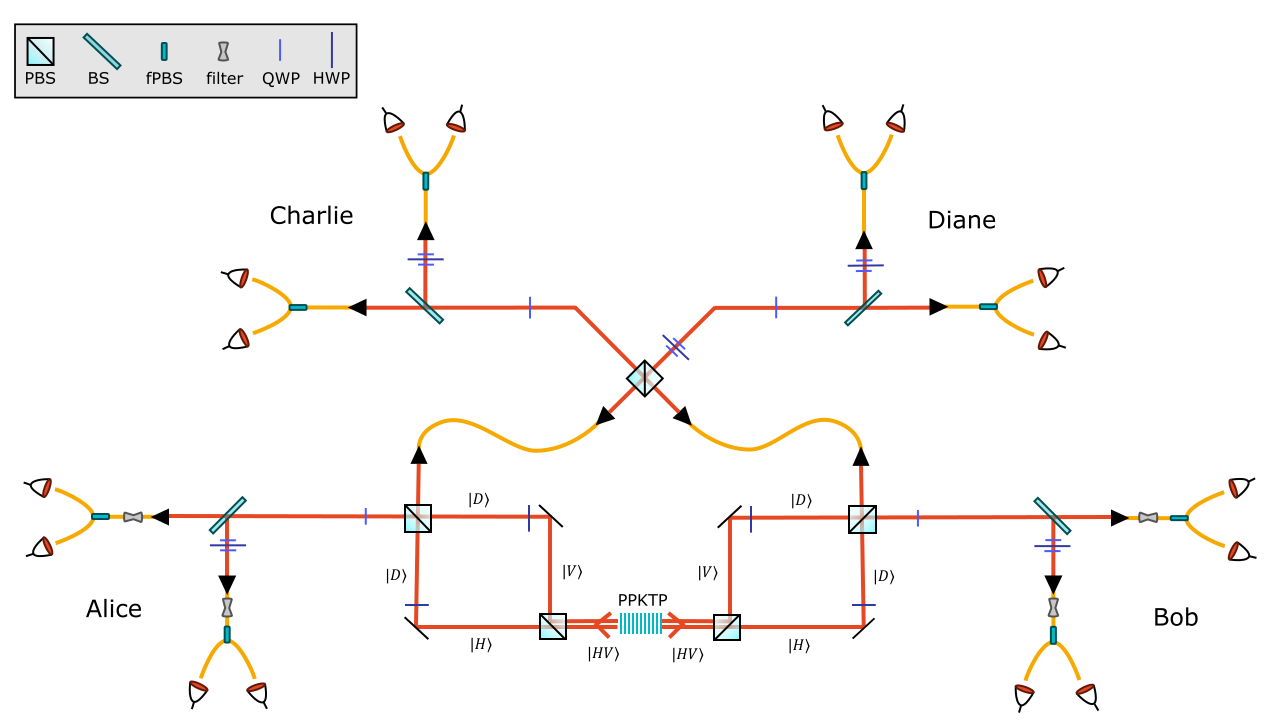}
\caption{\label{fig:fig_setup} Experimental setup (see text for details). PBS, polarizing beam splitter; BS, beam splitter; fPBS, fiber polarizing beam splitter; QWP, quarter waveplate, HWP, half waveplate; PPKTP, periodically-poled potassium titanyl phosphate.}
\end{figure*}

\textit{Experiment and results} -- As shown in Fig.~\ref{fig:fig_setup}, we generate two photon pairs using bidirectionally-pumped type-II spontaneous parametric down-conversion in periodically-poled potassium titanyl phosphate (PPKTP), tuned such that signal and idler are degenerate at 1543.7 nm. We use a ps-pulsed laser with a repetition rate of 76\,MHz, and maintain a photon pair per-pulse probability of $\approx.003$. The source produces relatively high purity states but we also filter each mode using DWDMs at 1543.73 nm (ITU C-Band Channel 42) to remove the SPDC side lobes and ensure high spectral purity \cite{Marcellino2024,Caspar2020}. Finally, photon detection is accomplished with SNSPDs (developed in-house), all with efficiencies above 80\% and dark count rates below 300 counts $s^{-1}$. We report a fourfold coincidence rate of approximately 0.3 events $s^{-1}$.


All four voters measure in two bases, each with two possible outcomes, corresponding to 16 total SNSPD detection channels. We realize the choice of measurement basis and subsequent rotations using a $50/50$ beamsplitter for each agent, followed by an identity or Hadamard gate and coupling into fiber. We set up 16 synchronous timestamp-streaming channels and process the timestamps on-the-fly with software developed in-house to avoid storing superfluous data. The first stage of processing involves a fourfold anti-coincidence (veto) filter on the timestamps from each voter using a pre-defined coincidence window of 1ns. This conditions further processing on each voter detecting exactly one event, effectively reducing spurious contributions due to detector dark counts and multi-pair SPDC photons. We then check the remaining timestamps for fourfold coincidences with the same window while tracking the timestamp channel numbers, giving us a list of fourfold events tagged to specific combinations of detector outcomes from each voter (this corresponds to the information available to the Verifier each round). Finally, this list is passed on to perform verification or determine the vote.  

We conduct quantum state tomography on the fourfold postselected state prior to the 50/50 beamsplitters. While full state tomography of a four-qubit density matrix normally requires 81 measurements for informational completeness, the low rank of the GHZ density matrix means we can make accurate estimates with far fewer measurements~\cite{Gross2010}. We perform 9 measurements on our state, each consisting of a tensor product of X, Y, or Z Pauli measurements, and reconstruct the density matrix using a purpose-built neural network~\cite{Koutny2022}. The reconstructed density matrix is shown in Fig. \ref{fig:fig_rho_real}. The results indicate a fidelity of $\approx 89\%$. As a check, we compute a lower bound on the fidelity by measuring a stabilizer of the GHZ state consisting of Pauli Z measurements on each qubit followed by Pauli X measurements~\cite{Somma2006}; the two methods agree to within error. 

We then execute the protocol, collecting fourfold coincidences for 5 hours before performing the verification test. The state passes verification $87\pm3\%$ of the time (as both voting and verification rely on the same principle -- that the state in question returns measurement results of determinate parity when measured in particular bases -- this is equivalently the likelihood of successful voting). 

We can attribute $\lesssim5.5\%$ of the failure rate to the imperfect fidelity of the initial state, as per the bound given in~\cite{Pappa2012}. The remainder comes as a consequence of our implementation strategy; by splitting the state four times using beamsplitters, we essentially have 16 sub-states after fourfold postselection, all of which might acquire slightly different phases that must be corrected for. In addition, the coincidence rate for each of these sub-states is reduced by a factor of $\frac{1}{16}$ with respect to the overall rate, making precise optimization challenging given low-frequency phase and polarization drifts. 


\textit{Conclusion} -- We have demonstrated a quantum electronic voting protocol with $\approx 87\%$ success probability based on a recent theoretical proposal~\cite{Centrone2022} that can guarantee anonymity for voters while requiring neither quantum memories nor assumptions on the behavior of the source. The latter advantage comes at the expense of increased experimental complexity, reflected in a success rate lower than what is in principle achievable given the fidelity of our state source. An alternate implementation that could avoid this particular failure mode uses fast polarization control for each voter, such that the decision of which basis to measure in and the corresponding rotations can be made after the state has left the source but before arriving to the voters (in practice, between pump laser pulses). This would eliminate the need for beamsplitters, as well as the associated rate reduction and phase alignment issues, but has the side effect of limiting the achievable distribution rate to the bandwidth of the polarization controller and relevant electronics. Scaling up to useful numbers of voters will require, first and foremost, practical sources of on-demand single photons; here, a promising candidate is semiconductor quantum dots~\cite{Senellart2017,Ding:2025}, which can be operated at rates reaching GHz~\cite{Tomm:2021}. To take advantage of these rates, an implementation like ours, using exclusively passive components, could then be preferred. 


\textit{Note} -- This work was conducted in parallel with an independent implementation of the original protocol by Nicolas Laurent-Puig, Matilde Baroni, and Eleni Diamanti at Sorbonne Universit\'{e}, and Federico Centrone at ICFO.

\textit{Acknowledgments} -- We thank Leili Esmaeilifar for useful discussions and Towsif Taher for assistance with SNSPDs. 

This work was supported by the Swiss State Secretariat for Research and Innovation (SERI) (Contract No. UeM019-3). M.W. is supported by the Swiss National Science Foundation through Ambizione Grant No. PZ00P2$\_$216153.

\medskip
\appendix
\section{Proof of anonymity for modified protocol}
In the original protocol, the Verifier chooses the measurement basis for each agent. In the modified protocol, the bases are instead chosen at random, and reported afterwards to the Verifier. One effect of this change is that the protocol can no longer certify that $\ket{\Phi_{0}^n}$ is shared between all parties. Indeed, the honest parties can no longer infer anything about the part of the state held by the dishonest parties. The honest parties can, however, make claims about the part of the state that they hold, and these will be sufficient to ensure that the protocol preserves anonymity i.e. the identity of the voter remains secret. 

We assume $k \ge 3$ honest parties and $n-k$ dishonest parties. The dishonest parties can collaborate both with each other and with the source; for example, they can choose which measurement bases and results they report to the Verifier as a function of the state that the source distributed on the associated round. 

We define the state $\ket{\Psi_0^k}$ as the result of applying 1 mod 4 Hadamard gates to a subset of the qubits of $\ket{\Phi_0^k}$ and $Z$ gates on the remaining qubits, and $\ket{\Psi_1^k}$ as the result of applying 3 mod 4 Hadamard gates to a subset of the qubits of $\ket{\Phi_0^k}$ and $Z$ gates on the remaining qubits. We begin by proving that, if the state passes verification with probability 1, then for every round the part of the distributed state held by the honest parties was one of $\{\pm \ket{\Phi_{0}^k},\pm \ket{\Phi_{1}^k},\pm \ket{\Psi_{0}^k},\pm \ket{\Psi_{1}^k}\}$.

\textit{Proof} -- Recall the verification criterion: $\frac{1}{2}H_p=\sum_iY_{i_p}$ mod 2, where $H_p$ is the total number of Hadamard measurements reported for event $p$, and must be even. Let $H_h$ be the number of Hadamard measurements made by the honest parties, and $H_d$ the number reported by the dishonest parties. In order to have $H_h + H_d$ even, $H_h$ and $H_d$ must have the same parity (in particular, the dishonest parties know that, if a round is not discarded, the parity of $H_h$ necessarily matches the parity of $H_d$). 

First, suppose $H_h$ and $H_d$ are both parity-even. Since $k \ge 3$, knowledge that $H_h$ is parity-even does not determine $\frac{1}{2}H_h$ mod 2. As such, the left side of the equality 
$\frac{1}{2}H_p=\sum_iY_{i_p}$ mod 2 can be either one or zero for any verification round. The dishonest parties have two actions available: they can choose to flip the parity of one or both of $\frac{1}{2}H_p$ mod 2 and $\sum_iY_{i_p}$ mod 2. In order to pass verification with probability 1, a malicious source must distribute a state to the honest parties such that the dishonest parties know with certainty which of these actions will satisfy the equality. This is only possible if each state ensures that the honest part of the verification test $\frac{1}{2}H_h=\sum_iY_{i_h}$ mod 2 either passes or fails with certainty. But from ~\cite{Pappa2012}, the only states that achieve this are $\pm \ket{\Phi_{0}^k}$, which cause the honest part of the test to pass with certainty, and $\pm \ket{\Phi_{1}^k}$, which cause it to fail with certainty. 

Now, suppose $H_h$ and $H_d$ are both parity-odd. Again, knowledge of the parity of $H_h$ does not determine $\frac{1}{2}(H_h + H_d)$ mod 2 for any $H_d$. The dishonest parties can then only pass verification with probability 1 if the source distributes a state to the honest parties such that the honest equality $\frac{1}{2}(H_h+1)=\sum_iY_{i_h}$ mod 2 is true or false with certainty. From ~\cite{Pappa2012} and the definition of $\ket{\Psi_{0}^k}$ and $\ket{\Psi_{1}^k}$, it is clear that the only states that achieve this are $\pm \ket{\Psi_{0}^k}$, which satisfy the equality with certainty, and $\pm \ket{\Psi_{1}^k}$, which falsify it with certainty.

In summary, the dishonest parties can adapt $H_d$ and $\sum_iY_{i_d}$ to pass verification with probability 1 if and only if the source distributes one of $\{\pm \ket{\Phi_{0}^k},\pm \ket{\Phi_{1}^k},\pm \ket{\Psi_{0}^k},\pm \ket{\Psi_{1}^k}\}$ on each round. This concludes the proof.

We now prove that voters remain anonymous even if the source is distributing $\pm \ket{\Phi_{0}^k}$, $\pm \ket{\Phi_{1}^k}$, $\pm \ket{\Psi_{0}^k}$, or $\pm \ket{\Psi_{1}^k}$ instead of $\ket{\Phi_{0}^n}$. First, note that honest parties behave identically with respect to all operations performed on the state whether voting or not, and no party can know in advance whether a round will be used for voting or verifying. As such, any distinguishing information could only come from the measurement results announced by the agents, where the voter might uniquely flip their measurement result bit. 

We have shown that the source can only pass verification with probability 1 if the dishonest parties report even $H_d$ on rounds where the source distributed $\pm \ket{\Phi_{0}^k}$ or $\pm \ket{\Phi_{1}^k}$ (in which case we must have $H_h$ even, else the round is discarded), and odd $H_d$ on rounds where the source distributed $\pm \ket{\Psi_{0}^k}$ or $\pm \ket{\Psi_{1}^k}$ (in which case we must have $H_h$ odd). Upon application of an even number of Hadamard gates to a subset of the constituent qubits, $\pm \ket{\Phi_{0}^k}$ and $\pm \ket{\Phi_{1}^k}$ go to $\pm \ket{\Phi_{0}^k}$ or $\pm \ket{\Phi_{1}^k}$, while upon application of an odd number of Hadamard gates, $\pm \ket{\Psi_{0}^k}$ and $\pm \ket{\Psi_{1}^k}$ also go to $\pm \ket{\Phi_{0}^k}$ or $\pm \ket{\Phi_{1}^k}$. 

We thus have that the honest state right before measurement in the computational basis is necessarily $\pm \ket{\Phi_{0}^k}$ or $\pm \ket{\Phi_{1}^k}$. But measurements on these states in the computational basis yield results evenly sampled from the spaces of parity-even or parity-odd $k$-bit strings, respectively. This property means that observing that the parity of the announced result string differs from expected (as when the voter flips their bit) gives no information about which bit was flipped. We conclude that our protocol preserves anonymity, at least in the limit where the state passes verification with probability 1. Given the properties of the embedded verification protocol and structural similarity to the original voting protocol, we expect our modified protocol to also have some finite error bound below which anonymity is preserved, but leave the calculation of this for future work. 

\bibliographystyle{myabbrvnat}
\bibliography{references}

\end{document}